\documentstyle[12pt,epsf,epsfig,wrapfig]{article}
\def\rf#1#2#3{{\bf #1}, #2 (19#3)}
\def\rft#1#2#3{{\bf #1}, #2 (20#3)}
\def\np{Nucl.\ Phys.\ }
\def\pr{Phys.\ Rev.\ }
\def\pl{Phys.\ Lett.\ }

\def\prl{Phys.\ Rev.\ Lett.\ }

\def\ppnp{Prog.\ Part.\ Nucl.\ Phys.\ }

\begin{document}
\begin{titlepage}
\rightline{\vbox{\halign{#\hfil \cr
ANL-HEP-PR-11-49 \cr
\today \cr}}}
\vspace{1.0cm}

\begin{center}
\Large                                                       
{\bf Studying Spin-Orbit Dynamics using Measurements of the Proton's Polarized Gluon Asymmetry} \\
\vspace{1.0cm}
\normalsize
Yevgeny Binder\footnote{Work supported in part by a Mulcahy Undergraduate Research Scholarship, Loyola University Chicago}\\{\it Quantum Code, Buffalo Grove, IL 60089}\\{\it and Physics Department, Loyola University, Chicago, IL 60626}\\
\medskip
Gordon P. Ramsey\footnote{Work supported by the U.S. Department of Energy,
Division of High Energy Physics, Contract DE-AC02-06CH11357. E-mail:
gpr@hep.anl.gov}\\{\it Physics Department, Loyola University, Chicago, IL
60626}\\{\it and High Energy Physics Division, Argonne National Lab, Argonne,
IL 60439} \\
\medskip
Dennis Sivers\\
{\it Portland Physics Institute, Portland, OR 97239}\\
{\it and Spin Physics Center, University of Michigan, Ann Arbor, MI 48103}
\end{center}

\begin{abstract}
Measurements involving the gluon spin density, $\Delta G(x,t)\equiv G_{++}(x,t)-G_{+-}(x,t)$, 
can play an important role in the quantitative understanding of proton structure. To 
demonstrate this, we show that the \underline{shape} of the gluon asymmetry, 
$A(x,t)\equiv \Delta G(x,t)/G(x,t)$, contains significant dynamical information about non-perturbative 
spin-orbit effects. It is instructive to use a separation $A(x,t)=A_0^{\epsilon}(x)+\epsilon(x,t)$, where $A_0^{\epsilon}(x)$ is an approximately scale invariant form that can be calculated within a given 
factorization prescription from the measured distributions, $\Delta q(x,t)$, $q(x,t)$ and $G(x,t)$. 
Applying this separation with the $J_z={1\over 2}$ sum rule provides a convenient way to 
determine the amount of orbital angular momentum generated by mechanisms associated with 
confinement and chiral dynamics. The results are consistent with alternate non-perturbative approaches
to the determination of orbital angular momentum in the proton. Our studies help to specify the accuracy 
that future measurements should achieve to constrain theoretical models for nucleon structure.
\end{abstract}

\setcounter{footnote}{0} 
\end{titlepage}

\section{Introduction}

Several experimental programs \cite{hermes,SMC2,star,PHENIX,compass3} have devised 
strategies aimed at providing a significant measure of the proton's spin weighted gluon density,
\begin{equation}
\Delta G(x,t)\equiv G_{++}(x,t)-G_{+-}(x,t),
\end{equation}
where $x$ is the Bjorken scaling variable and $t\equiv \log[\alpha_s(Q_0^2)]/\log[\alpha_s(Q^2)]$
is the $Q^2$ evolution variable. The interest in these measurements is often framed in terms of 
arguments using the $J_z={1\over 2}$ sum rule, \cite{bms}
\begin{equation}
J_z={1\over 2}\equiv {1\over 2}<\Delta \Sigma(t)>+<\Delta G(t)>+L_z(t), \label{JSR}
\end{equation}
where 
\begin{eqnarray}
<\Delta \Sigma(t)>&\equiv&\int^1_0\>dx\Delta q(x,t) \quad \mbox{and}
\nonumber \\
<\Delta G(t)>&\equiv&\int^1_0\>dx\Delta G(x,t)
\end{eqnarray}
are the projections of the spin carried by all quarks and the gluons on the helicity (or $z$)-axis, 
respectively. Also $L_z(t)$ is the net $z$-component of the orbital angular momentum of the 
constituents. In this analysis, we will not attempt to identify independent flavor components of $L_z(t)$ 
within the sum rule. Arguments concerning the content and assumptions involved in such separations 
have been presented by Jaffe and Manohar \cite{jm} and by Ji. \cite{jilz} Since these theoretical 
considerations do not impact the content of the analysis in sections 1 and 2, we will defer to the subject
of separating quark and gluon orbital angular momenta until section 3, where we discuss specific 
models for proton structure.

The $t$-evolution of the terms on the RHS of equation (\ref{JSR}) adds some content to this discussion. It 
was shown in \cite{bms} that the lowest order QCD evolution equation leads to 
${\partial\Sigma}/{\partial t}=0$ in the chiral factorization prescription, reflecting the underlying helicity 
conservation of QCD perturbation theory. It was also shown that for a large range of boundary 
conditions, ${\partial\Delta G}/{\partial t}>0$, so that the sum rule (\ref{JSR}) would seem to lead to the conclusion that
\begin{eqnarray} 
\lim_{t \to\infty} <\Delta G(t)>\to +\infty \\ 
\lim_{t \to\infty} \L_z(t)\to -\infty. \label{DGL}
\end{eqnarray} 
The reconciliation of this result within the framework of perturbative QCD was achieved by 
Ratcliffe \cite{ratcliffe}, who showed that the structure of the DGLAP evolution equations \cite{dglap} is 
based upon an iteration of perturbative processes in which a change of $\Delta G=\pm 1$ is balanced by 
a corresponding change of $L_z=\mp 1$. The chiral structure of QCD therefore implies that a 
helicity-sensitive probe of the proton's gluon density uncovers $Q^2$ dependence from Ratcliffe 
resolution structures that can be considered perturbatively generated P-wave color correlations of the 
composite system. Higher order evolution \cite{dglap} makes small modifications to Ratcliffe's result without diminishing its basic content. Recent papers by Bakker, Leader and Trueman \cite{blt} and by
Chen and Ji, \cite{chenji} have clarified the understanding of angular momentum sum rules by 
dealing with some of the complications associated with the non-local properties of orbital angular momentum. 

Recent DIS experiments \cite{hermes2,compass2} have significantly lowered the measurement errors 
of the quark spin contribution ($\Delta \Sigma$) to equation (\ref{JSR}). The COMPASS collaboration 
analysis quotes a result
\begin{equation}
<\Delta \Sigma>=0.30\pm 0.01 (\mbox{stat}) \pm 0.02 (\mbox{evol}) ,\qquad \mbox{all data} \label{DSigC}
\end{equation}
while the HERMES collaboration analysis quotes a result
\begin{equation}
<\Delta \Sigma>=0.330\pm 0.025 (\mbox{exp}) \pm 0.011 (\mbox{th})\pm 0.028 (\mbox{evol}),\qquad \mbox{all data} \label{DSigH}.
\end{equation}
These values can be used with the $J_z={1\over 2}$ sum rule to evaluate the impact of existing and 
potential gluon asymmetry measurements. 

Recent experimental results sensitive to $\Delta G(x,t)$ and the gluon asymmetry, 
$A(x,t)\equiv \Delta G(x,t)/G(x,t)$ have provided important new information. Although the analysis of 
these experiments is limited in sensitivity and range of $<x>$, the results suggest the possibility that 
at some small $x$ and $Q_0^2$, $\Delta G(x,t_0) \le 0.$ Understanding the shape of $\Delta G(x,t)$
for the whole range $x\in (0,1)$ is important, since the asymptotic result 
\begin{equation}
\lim_{t \to\infty} \alpha_s(t) <\Delta G(t)> \> = \mbox{constant} \label{asympt}
\end{equation}
follows from the full QCD evolution equations, where the constant could be positive, negative or zero
as $t\to \infty$. These measurements suggest that phenomenological conclusions described in 
equations (\ref{DGL}) may be in error and the asymptotic color structure of the proton is quite different 
from what has been previously supposed. If the constant in equation (\ref{asympt}) is zero, then
$<\Delta G(t)>$ and $L_z(t)$ should asymptotically approach constants, while if it is negative, then the 
signs of equations (\ref{DGL}) should be switched. Thus, experimental evidence on $\Delta G(x,t_0)$
for a limited range of $x$ and $t_0$ must be combined with an extrapolation in order to specify the
nature of $<\Delta G(t)>$ and $L_z(t)$ at large $t$. The experimental results alone are not conclusive. 
The specific approach in this paper helps to illuminate these possibilities and to fix onto crucial 
experimental results.

From the discussion above, it is instructive to write the polarized gluon asymmetry using the decomposition 
\begin{equation}
A(x,t)\equiv \Delta G(x,t)/G(x,t)=A_0^{\epsilon}(x)+\epsilon(x,t), \label{Adef}
\end{equation}
where $A_0^{\epsilon}(x)$ is dependent only upon $x$, calculable in PQCD by the using definition 
\begin{equation}
A_0^{\epsilon}(x)\equiv \Bigl[({{\partial \Delta G(x,t)}\over {\partial t}})/
({{\partial G(x,t)}\over {\partial t}})\Bigr]. \label{A0def}
\end{equation}
The numerator and denominator on the right side of equation (\ref{A0def}) are calculable from the
DGLAP equations and each depends strictly upon $x$ via the respective convolutions. \cite{dglap}
The small correction, $\epsilon(x,t)$ describes shape-dependent differences in the evolutions of $G(x,t)$ and $\Delta G(x,t)$ at leading order (LO) in QCD perturbation theory. Differentiating $A(x,t)$ with respect to $t$ in equation (\ref{Adef}) and using the scale-invariance of $A_0^{\epsilon}(x)$, it follows that 
$$\frac{\partial}{\partial t}[\epsilon(x,t)\>G(x,t)]=0.$$
The expression (\ref{Adef}) for $A(x,t)$ at some initial $t=t_0$ leads to an equivalent decomposition for 
$\Delta G(x,t)$ in the form
\begin{equation}
\Delta G(x,t_0)=A_0^{\epsilon}(x)\cdot G(x,t_0)+\Delta g_{\epsilon}(x) \label{DGdef}
\end{equation}
where the ``polarized gluon excess'', $\Delta g_{\epsilon}(x)$, is given by
\begin{equation}
\Delta g_{\epsilon}(x)=\epsilon(x,t)\cdot G(x,t) \label{DGe}
\end{equation}
and is $t-$independent. This provides a boundary condition for the partial differential equation 
(\ref{A0def}) that defines $A_0^{\epsilon}(x)$ and can be used to characterize possible different shapes 
for $A(x,t)$ in equation (\ref{Adef}). This boundary condition for the partial differential equation 
(\ref{A0def}) occurs at an unphysical region in that the decomposition in (\ref{DGdef}) cannot be valid 
when $G(x,t_0)=0$. In practice, this means that there are nontrivial constraints on the magnitude of
$\Delta g_{\epsilon}(x)$. We will discuss these constraints when we consider explicit solutions to
equation (\ref{A0def}).

In this paper, we will neglect the $t$-dependence of $\Delta \Sigma (x,t)$ at LO and concentrate on an 
alternate approach to characterizing the consequences of the $t$-dependence of $\Delta G(x,t)$.
Using the current  data for $\Delta \Sigma$ as input, we combine equations 
(\ref{JSR}), (\ref{DSigC}) and (\ref{DSigH}) to write
\begin{equation}
L_z(t)+<\Delta G(t)>\approx {1\over 2}(1-<\Delta \Sigma>)\approx 0.34\pm 0.02 \label{Leps}
\end{equation}
in a chiral factorization prescription. The quoted error is entirely due to the data uncertainties in
equations (\ref{DSigC}) and (\ref{DSigH}). The asymmetry $A(x,t)$ is parameterized as outlined in 
Section 2 and the corresponding polarized gluon obtained from equation (\ref{Adef}) is substituted into 
equation (\ref{Leps}) to determine a value for the orbital angular momentum, $L_z(t)$. Our approach to 
the study of $\Delta G(x,t)$ is largely complementary to the usual global analysis determination discussed, for example, by Hirai and Kumano \cite{acc} and others \cite{bmn,bb} where the main 
input for $\Delta G(x,t)$ involves measurement of the scaling violations for $\Delta q(x,t)$. Since it is 
highly unlikely that future experiments sensitive to $\Delta G(x,t)$ will determine this distribution with an 
accuracy similar to that found in the determination of $\Delta q$ or to that of $G(x,t)$ and $q(x,t)$, our 
method takes into account the similarities between the evolution of $\Delta G(x,t)$ and $G(x,t)$ to 
provide the necessary extrapolations to all $x$. Using knowledge of $\Delta q(x,t)$ along with $q(x,t)$ 
and $G(x,t)$ data sensitive to values of $\Delta G(x,t)$ in limited regions  of $x$ and $t$ can then be 
used efficiently to obtain valid estimates for $L_z(t)$. In particular, as long as $G(x,t_0)$ is large enough 
at small scales, $t_0$, associated with confinement and chiral dynamics, $L_z(t)$ can also be 
extrapolated to provide a measure of spin-orbit dynamics at such scales. 

The possibility that there could exist a scale-invariant form for the gluon polarization asymmetry, $A(x,t)$, 
was first considered by Einhorn \cite{ein} on the basis of numerical studies of the LO DGLAP equations. \cite{dglap} The work of reference \cite{cs} explores Einhorn's suggestion by showing that a scale-invariant form for $A_0^{\epsilon}(x)$ is a feature of the Close-Sivers Bremsstrahlung model. 
Scale-invariance occurs in that approach because the perturbative diagrams which determine the 
$t$-evolution of the gluon distributions are the same as those that distribute the spin information to the 
gluons at LO. The papers of references \cite{gpr}-\cite{gpr3} have significantly expanded the 
phenomenological study of the information contained in the shape of $A(x,t)$ by introducing the 
connection to the $J_z=1/2$ sum rule and by considering the possibility of small fluctuations around the 
scale-invariant form described in equations (\ref{Adef}), (\ref{DGdef}) and (\ref{DGe}). 

The remainder of this paper is organized as follows. 
In Section 2 we discuss how equation (\ref{A0def}) leads to a non-linear equation for 
$A_0^{\epsilon}(x)$ based on the DGLAP equations. We then generate parameterizations of 
$A_0^{\epsilon}(x)$ in leading-order (LO) and outline the next-to-leading order (NLO) calculation. These 
parameterizations can give a meaningful description of $A(x,t)$ or $\Delta G(x,t)$ with suitable constraints 
on $\epsilon(x,t)$ or $\Delta g_{\epsilon}(x)$ respectively. When the constraints are satisfied, a complete 
description of $A_0^{\epsilon}(x)$ is obtained from boundary conditions combined with existing 
measurements of the singlet quark and unpolarized gluon densities, $q(x,t)$, $\Delta q(x,t)$ and 
$G(x,t)$. Section 3 presents numerical determination of $A_0^{\epsilon}(x)$ within these constraints and 
the observation that preliminary data on $\Delta G(x,t)$ from several  measurements are within these 
bounds. This suggests that future measurements of $\Delta G(x,t)$ or $A(x,t)$ with improved accuracy 
can be used in conjunction with our analysis to determine a low energy non-perturbative component of 
orbital angular momentum. Section IV includes a brief discussion of the importance of understanding the 
constituent orbital angular momentum in a composite system and how the $J_z={1\over 2}$ sum rule 
and measurements sensitive to $A(x,t)$ can provide important constraints on dynamical models for 
proton structure. 

\section{The Scale-Invariant Gluon Asymmetry $A_0^{\epsilon}(x)$}

\subsection{Numerical approach to the gluon asymmetry}

The solution of equations based on equations (\ref{Adef}) and (\ref{A0def}) was proposed by Ramsey 
and Sivers [21-23]. The calculation of the asymmetry provided a method to determine $\Delta G$ 
without theoretical biases on its shape. This was followed by analysis of the relation between the 
$\Delta g_{\epsilon}$ parameterizations and the corresponding range of possible $L_z$ \cite{gpr1}. Since 
new data for the asymmetry have been made available, this method has allowed us to refine the range 
of possible $<\Delta G>$ and $L_z$ consistent with this data \cite{gpr2,gpr3}. The culmination of this 
work is presented in this paper. 

In this section, we create a numerical mechanism which allows phenomenological determination of the 
non-perturbative quantity $L_z$ from the experimental measurements of $\Delta G$ and 
${{\Delta G}\over G}$. This is done without an a priori assumption for the polarized glue. We will outline a 
numerical approach to calculate the asymmetry-$L_z$ connection in order to provide a 
self-consistency check to this approach and provide a range of possible values for the orbital angular 
momentum.

The perturbative component of the polarized gluon asymmetry $A_0^{\epsilon}(x)$ can be calculated from a parameterization of the correction $\Delta g_{\epsilon}(x)$ by inserting equation (\ref{DGdef}) into 
the expression (\ref{A0def}) for $A_0^{\epsilon}(x)$. Then equation (\ref{A0def}) can then be solved 
using $\partial \Delta G/\partial t$ and $\partial G/\partial t$ given by DGLAP evolution. In kinematic 
regions where the DGLAP evolution equations are valid, equation (\ref{A0def}) allows one to generate 
$A_0^{\epsilon}(x)$ by 
\begin{equation}
A_0^{\epsilon}(x)=\Biggl[{{\Delta P_{gq} \otimes \Delta q+\Delta P_{gg}\otimes (\Delta G)}\over {P_{gq} \otimes q+P_{gg}\otimes G}}\Biggr], \label{A0calc}
\end{equation}
where the convolution is defined as $P(x)\otimes Q(x)\equiv \int_0^1 \frac{dy}{y} P(y)Q(x/y)$. Then 
equations (\ref{A0def}) and (\ref{DGdef}) can be used to generate the corresponding asymmetry using 
the DGLAP equations via the iterative equation
\begin{equation}
A_0^{\epsilon}(x)=\Biggl[{{\Delta P_{Gq} \otimes \Delta q+\Delta P_{GG}\otimes [A_0^{\epsilon}
\cdot G+\Delta g_{\epsilon}]}\over {P_{Gq} \otimes q+P_{GG}\otimes G}}\Biggr]. \label{A0calc2}
\end{equation}
Since the $A_0^{\epsilon}(x)$ term occurs on both sides of the equation, we parametrize 
$A_0^{\epsilon}(x)$ and $\Delta g_{\epsilon}(x)$ subject to theoretical constraints and determine the
coefficients of the parameterizations that satisfy equation (\ref{A0calc2}). For each fixed parameterization 
of $\Delta g_{\epsilon}(x)$, the resulting asymmetry $A(x,t)$ and corresponding $\Delta G$ from 
equation (\ref{DGdef}) can then be checked for positivity with the corresponding unpolarized gluon at 
LO. The various parametrizations of $\Delta g_{\epsilon}(x)$ were chosen to have integrals over all
$x$ between $-0.5$ and $0.5$. The particular parametrization in the second line of Table 1 was
selected to change sign, consistent with instanton models. This possibility was discussed by the HERMES collaboration and has not been ruled out by data. \cite{hermes,hermes2,acc,lss}
The process to determine the NLO corrections to the asymmetry is similar to that of LO with the 
appropriate corrections to the splitting functions and evolution parameters. This will be addressed 
shortly. 

The calculation of $A_0^{\epsilon}(x)$ in equation (\ref{A0calc2}) has been done using the 
CTEQ5 \cite{cteq} and MRST \cite{mrst} unpolarized parton distributions for comparison and the 
polarized quark distributions from reference \cite{ggr}. All are evaluated at $Q_0^2=1$ GeV$^2$ for the 
LO distributions. The CTEQ5 unpolarized distributions are given by \\
\begin{eqnarray}
xu_v(x)&=&0.9783 x^{0.4942}(1-x)^{3.3705}(1+10.0012 x^{0.8571}) \nonumber \\
xd_v(x)&=&0.5959 x^{0.4942}(1-x)^{4.2785}(1+8.4187 x^{0.7867}) \nonumber \\ 
xS_{total}(x)&=&1.2214 x^{0.0877}(1-x)^{7.7482}(1+3.389 x) \nonumber \\
xG(x)&=&3.3862 x^{0.261}(1-x)^{3.4795}(1-0.9653 x). \label{CTEQ5}
\end{eqnarray}
For this set, $\int_0^1 xG(x)\>dx=0.28$, representing the momentum carried by the gluons.
The MRST LO unpolarized distributions are \\
\begin{eqnarray}
xu_v(x)&=&0.6051 x^{0.4089}(1-x)^{3.395}(1+2.078 \surd{x}+14.56 x) \nonumber \\
xd_v(x)&=&0.05811 x^{0.2882}(1-x)^{3.874}(1+34.69 \surd{x}+28.96 x) \nonumber \\
xS_{total}(x)&=&0.2004 x^{-0.2712}(1-x)^{3.808}(1+2.283 \surd{x}+20.69 x) \nonumber \\
xG(x)&=&64.57 x^{0.9171}(1-x)^{6.587}(1-3.168 \surd{x}+3.251 x). \label{MRST}
\end{eqnarray}
For the MRST set, $\int_0^1 xG(x)\>dx=0.35$, for comparison to the CTEQ5 set. In each case the total 
quark contribution is then given by:
\begin{equation}
xq_{total}(x)=xu_v(x)+xd_v(x)+xS_{total}(x).
\end{equation}

The polarized distributions given by GGR \cite{ggr} in terms of the unpolarized ones are \\
\begin{eqnarray}
x\Delta u_v(x)&=&[1+0.25(1-x)^2/\surd x]^{-1} (xu_v(x)-2xd_v(x)/3) \nonumber \\
x\Delta d_v(x)&=&[1+0.25(1-x)^2/\surd x]^{-1} (-xd_v(x)/3) \\
\Delta S_{total}(x)&=&(-2.36+2.66\surd x)\>xS_{total}(x), \nonumber 
\end{eqnarray}
where
\begin{equation}
x\Delta \Sigma (x)=x\Delta u_v(x)+x\Delta d_v(x)+x\Delta S_{total}(x). \label{polq}
\end{equation}

Although, in principle, we are sensitive to the shape of the unpolarized gluon distribution at $Q_0^2$,
the CTEQ and MRST only differ at small $x<0.1$ and the effect on our results is minimal.
For both unpolarized CTEQ and MRST distributions used in equation (\ref{A0calc2}), the term in the 
denominator vanishes at a critical Bjorken-$x$ value of approximately $x_c\approx 0.20\to 0.30$ at LO. 
To avoid this numerical problem, we take the denominator and terms with the asymmetry to one side, 
giving
\begin{equation}
A_0^{\epsilon} \Bigl[P_{gq}^{LO}\otimes q+P_{gg}^{LO}\otimes G\Bigr]-\Delta P_{gg}^
{LO}\otimes (A_0^{\epsilon}\cdot G)=\Bigl[\Delta P_{gq}^{LO}\otimes \Delta q+\Delta P_{gg}^
{LO}\otimes (\Delta g_{\epsilon})\Bigr]. \label{A0calc3}
\end{equation}
This mitigates the problem of the point discontinuity in equation (\ref{A0calc2}) and makes the 
determination of $A_0^{\epsilon}(x)$ straightforward. Since the asymmetry appears in both terms on the 
left hand side of the equation, a numerical trial-and-error technique must be used. 

To establish an initial reference point for the solution of equation (\ref{A0calc3}), practical constraints for the asymmetry at LO should include:
\begin{itemize}
\item strong positivity: $|A(x,t_0)| \le 1$ for all $0\le x\le 1$, and
\item endpoint values: $A(0,t_0)=0$ and $A(1,t_0)=1$.
\end{itemize}
Simple positivity for the gluon asymmetry at LO requires $|A(x,t)|\le 1$. Since the individual terms on the 
right side of equation (\ref{DGdef}) do not represent separate observables, the calculated 
$A_0^{\epsilon}(x)$ need not satisfy these criteria, but the full calculated asymmetry, $A(x,t)$ must. 
As a practical matter, the strong positivity constraint enforces the restriction that $\Delta g_{\epsilon}(x)$ is small and provides a tight restraint on the range of $\Delta g_{\epsilon}(x)$ for which stable solutions 
can be found. The constraint on endpoint values enforces the Brodsky-Farrar constituent counting rules as $x\to 1$ and the requirement that $\Delta G(x=0,t)$ vanishes. In all solutions, the convolutions in
(\ref{A0calc3}) are dominated in the proton by the valence up quark as $x\to 1$.

To investigate possible asymmetry solutions, we start with an $A_0^{\epsilon}(x)$ parameterization in the 
form
\begin{equation}
A_0^{\epsilon}(x)\equiv Ax^{\alpha}-(B-1)x^{\beta}+(B-A)x^{\beta+1}, \label{A0epar}
\end{equation}
which automatically satisfies the endpoint constraints provided the exponents $\alpha$ and $\beta$ are 
positive. This parameterization includes substantial flexibility in adjusting shapes while keeping the 
number of free parameters to a minimum. Using the CTEQ and MRST unpolarized distributions and the GGR polarized distributions given in equations (\ref{CTEQ5}) through (\ref{polq}) for the calculation, we 
find that only parameterizations of $\Delta g_{\epsilon}$ that conform to the condition
\begin{equation}
|<\Delta g_{\epsilon}>| \le 0.25. \label{dgebnd}
\end{equation}
lead to solutions of equations (\ref{A0calc2}) and (\ref{A0calc3}) that satisfy the positivity constraint. 
A representative sample of the parameterizations whose asymmetries and corresponding polarized 
gluon distribution satisfy all the constraints are summarized in Table 1. These solutions give integrals 
$<\Delta G>$ ranging from about -0.09 to 0.59 at these small $Q_0^2=1$ GeV$^2$ values. Line two of the table contains the only non-monotonic parameterizations of $\Delta g_{\epsilon}$ that has been 
included in this sample. It should be noted that although some of the magnitudes of the integrated
gluon distributions, $<\Delta G>$, are less than those of $<\Delta g_{\epsilon}>$, they occur for the
parameterizations of $\Delta g_{\epsilon}$ whose integrals are negative or zero. Since this term is
added to the overall polarized gluon distribution, the results are consistent with the model.

\begin{table}[htdp]
\caption{Gluon Asymmetry Parameters at $Q_0^2=1$ GeV$^2$}
\begin{center}
\begin{tabular}{||c|c|c|c||}
\hline
$\Delta g_{\epsilon}$ & $<\Delta g_{\epsilon}>$ & $A_0^{\epsilon}$ & $<\Delta G>$ \\
\hline
\hline
$0$ & $0$ & $3x^{1.5}-3x^{2.2}+x^{3.2}$ & 0.28 \\
\hline
$x(x-0.25)(1-x)^5$ & $0$ & $1.75x^{1.9}-5x^{2.4}+4.25x^{3.4}$ & -0.01 \\
\hline
$2(1-x)^7$ & $0.25$ & $4x^{1.6}-4x^{2.1}+x^{3.1}$ & 0.51 \\
\hline
$-2(1-x)^7$ & $-0.25$ & $1.75x^{1.1}-1.5x^{2.1}+0.75x^{3.1}$ & 0.18 \\
\hline
$18x(1-x)^7$ & $0.25$ & $2x^{1.8}-x^{2.5}$ & 0.39 \\
\hline
$-18x(1-x)^7$ & $-0.25$ & $2.25x^{1.7}-x^2-0.25x^3$ & -0.09 \\
\hline
$-90x^2(1-x)^7$ & $-0.25$ & $3.5x^{1.3}-4.5x^{2.2}+2x^{3.2}$ & 0.25 \\
\hline
$9x(1-x)^7$ & $0.125$ & $3.75x^{1.4}-3x^{1.6}+0.25x^{2.6}$ & 0.40 \\
\hline
$-9x(1-x)^7$ & $-0.125$ & $3.25x^{1.4}-3.75x^{2.2}+1.5x^{3.2}$ & 0.25 \\
\hline
$4.5x(1-x)^7$ & $0.0625$ & $2x^{0.9}-1.5x^{1.2}+0.5x^{2.2}$ & 0.59 \\
\hline
$-4.5x(1-x)^7$ & $-0.0625$ & $2.25x^{1.1}-2.25x^{1.9}+x^{2.9}$ & 0.43 \\
\hline
\end{tabular}
\end{center}
\label{default}
\end{table}

The solutions for $A^{\epsilon}_0$ listed in Table 1 represent the best fit coefficients of equation 
(\ref{A0epar}) subject to roundoff error in solving equation (\ref{A0calc3}). These parametrizations
of $A^{\epsilon}_0$ are used to determine forms for $\Delta G(x,t)$ in equation (\ref{DGdef}) that can be tested experimentally in processes at large $t$ for various ranges of $x$. The scale invariant nature 
of $\Delta g_{\epsilon}(x)$ implies that ${\partial \Delta g_{\epsilon}}/{\partial t}=0$, so that the accuracy 
by which we can extract information on $L_z$ can be estimated. Expanding 
$\Delta g_{\epsilon}=\epsilon(x,t)\cdot G(x,t)$ in $t$:
\begin{equation} 
\epsilon(x,t)\cdot G(x,t)=\epsilon(x,t_0)\cdot G(x,t_0)+{{\partial(\epsilon(x,t)\cdot G(x,t))}\over {\partial t}}(t-t_0)+{{\partial^2(\epsilon(x,t)\cdot G(x,t))}\over {\partial t^2}}(t-t_0)^2+\cdots. \label{dgeexp}
\end{equation}
The second term on the right side vanishes identically and at LO and the second order term in 
equation (\ref{dgeexp}) can be neglected, since second order terms in $t$ may arise only at NLO and 
higher. Thus, at LO we can write
\begin{equation}
\Delta G(x,t)-\Delta G(x,t_0)\approx A_0^{\epsilon}(x)\cdot [G(x,t)-G(x,t_0)]+\mathcal{O}(t-t_0)^2. 
\label{DDG}
\end{equation}
Thus, the polarized gluon distribution can be parametrized accurately at LO. All of the perturbatively 
generated orbital angular momentum can be extracted from the $J_z=1/2$ sum rule within parton distribution uncertainties to this order. At LO, 
$\Delta \Sigma(x)$ is independent of $t$, so the amount of orbital angular momentum generated by 
Ratcliffe resolution structures in the range of scales $(t,t_0)$ can be given by:
\begin{equation}
L_z(t)-L_z(t_0)=-\int_0^1 dx\>A_0^{\epsilon}\cdot [G(x,t)-G(x,t_0)]+\mathcal{O}(t-t_0)^2. \label{Lbrems}
\end{equation}
Solving the nonlinear equation for $A_0^{\epsilon}(x)$ then leads to what we call the $\underline{Ratcliffe-subtracted}$ version of the $J_z={1\over 2}$ sum rule, where the perturbatively generated 
orbital angular momentum and the perturbative evolution of $\Delta G(x,t)$ are subtracted. The 
separation allows for the measurement of $\Delta G(t)$ to be associated with a non-perturbative 
component of orbital angular momentum generated by the effects of chiral symmetry and confinement 
on proton structure. Using equations (\ref{Leps}) and (\ref{Lbrems}) we can write
\begin{equation}
L_z(t_0)=0.34\pm 0.02 -<\Delta G(t)> + <A_0^{\epsilon}\cdot [G(x,t)-G(x,t_0)]>+\mathcal{O}(t-t_0)^2. \label{Lt0}
\end{equation}
so that measurements sensitive to $\Delta G(x,t)$ at some large value of $t$ can be used to estimate
the amount of constituent orbital angular momentum present at smaller values of $t$ in the perturbative region. For this study, we will choose $t_0\equiv (t=0)$ to correspond to $Q^2=1\>GeV^2$. This is a 
region where the DGLAP evolution is valid and processes evolved by BFKL evolution are not crucial to this study. It is also the kinematic region where chiral symmetry is broken. All parton distributions are 
evolved to this value. 

Our preliminary studies into the numerical solutions to equation (\ref{A0calc3}), as indicated in Table 1, 
demonstrate that such solutions are sensitive to $\underline{both}$ the shape of $\Delta g_{\epsilon}(x)$
$\underline{and}$ its normalization, $\mid <\Delta g_{\epsilon}>\mid$. This dual sensitivity arises 
because the region in $x$ where the denominator of equation (\ref{A0calc2}) is small,
$x\in (0.24,0.30)$, contributes significantly to $A_0^{\epsilon}(x)$ for all $x\le 0.30$. The nonlinear nature of this equation makes the solution sensitive to the convolution term, 
$\Delta P_{gg}^{LO}\otimes \Delta g_{\epsilon}$ on the right side of (\ref{A0calc3}). As indicated in the
discussion above, the requirement in (\ref{dgebnd}) limits the impact of the normalization. To further 
study the numerical shape of $\Delta g_{\epsilon}(x)$, we have separately examined solutions to
(\ref{A0calc3}) where the support of $\Delta g_{\epsilon}(x)$ was restricted to the large$-x$ region,
$x\ge 0.30$, or to the small$-x$ region, $x\le 0.20$. 

These studies indicated that $\Delta g_{\epsilon}(x)$ with support in the large$-x$ region, the condition
$\Delta g_{\epsilon}(x)\ll \Delta q(x,t)$ is required uniformly in $x$ and $t$ for solutions to be consistent
with the positivity condition and with the expectation that $\Delta q(x,t)/q(x,t)\to 1$ as $x\to 1$. When this
condition is valid, the impact of $\Delta g_{\epsilon}(x)$ on $A_0^{\epsilon}(x)$ can be ignored 
compared with the errors in $\Delta q(x,t)$. Restrictions on parameterizations of $\Delta g_{\epsilon}(x)$ 
in the small$-x$ region were found to be much less stringent and and as long as the normalization 
requirement (\ref{dgebnd}) was valid, variations in the shape in the region did not impede the ability to 
find stable solutions. These studies demonstrate that our approach of parametrizing the shape of 
$A(x,t)$ can provide an independent, complementary approach to the global fits \cite{acc}-\cite{bb} 
directed at the problem of examining the impact of data on $\Delta G(x,t)$ for the determination of 
$<\Delta G(t)>$. In particular,  we can formulate consistent parameterizations in which the asymmetry, 
$A(x,t)$, changes sign.

The phenomenological application of (\ref{Lt0}) to the determination of $L_z(t_0=0)$ implies an
understanding of the errors involved in determining the components on the right hand side of
(\ref{Lt0}). In addition to the uncertainty associated with measurements of $<\Delta \Sigma>$, which is
known explicitly, there are errors associated with
\begin{enumerate}
\item the determination of $<\Delta G(t)>$ from experiments with limited acceptance,
\item the numerical determination of $A_0^{\epsilon}(x)$ from solutions to (\ref{A0calc3}) and
\item the continuation to $t=0$ using $<A_0^{\epsilon}[G(t)-G(0)]>$ within this framework.
\end{enumerate}

We will address the issues in the first two contributions in more detail in section 3. The errors implied
by the third point can, in principle, be minimized by solving (\ref{A0calc}) at NLO in QCD perturbation
theory instead of the LO solutions described by (\ref{A0calc3}). This discussion follows here.

\subsection{Next-to-leading Order Asymmetry}

While we have formulated the arguments above in terms of the solutions to equation (\ref{A0calc}) using
the LO DGLAP equations, it is appropriate at this point to insert a few comments on the application of 
solutions to the next-to-leading order set of equations 
\begin{eqnarray}
\Delta G^{NLO}(x,t)&=&A_0^{\epsilon,NLO}(x,t_1)\cdot G^{NLO}(x,t)+\Delta g_{\epsilon}(x) \nonumber \\
A_0^{\epsilon,NLO}(x,t_1)&\equiv &\Bigl[({{\partial \Delta G^{NLO}(x,t)}\over {\partial t}})\mid_{t=t_1}/
({{\partial G^{NLO}(x,t)}\over {\partial t}})\mid_{t=t_1}\Bigr] \label{DGNLO}
\end{eqnarray}
to the analysis of data sensitive to the polarized gluon asymmetry. Here $t_1$ is a reference $t$ at
NLO. Such phenomenological analyses have traditionally been done in the framework of LO 
perturbation theory. The non-linear equation analogous to (\ref{A0calc2}) that must be solved 
numerically then can be written in the form 
\begin{eqnarray}
A_0^{\epsilon,NLO}(x,t_1)= {{\Bigl[\Delta P_{Gq}^{NLO} \otimes \Delta q^{NLO}(x,t) + \Delta P_{GG}^{NLO} \otimes [A_0^{\epsilon,NLO}(x,t_1)\cdot G^{NLO}(x,t)+\Delta g_{\epsilon}(x)]\Bigr]}\over {\bigl[P_{Gq}^{NLO} \otimes q^{NLO}(x,t) + P_{GG}^{NLO} \otimes G^{NLO}(x,t)\bigr]}}
\end{eqnarray}
where it is specified that the convolutions on the right side are performed with the appropriate distribution functions: $q^{NLO}(x,t)$, $G^{NLO}(x,t)$, $\Delta q^{NLO}(x,t)$, $\Delta G^{NLO}(x,t)$
and the NLO splitting functions all evaluated at $t=t_1$. This specification acknowledges that the
splitting functions evaluated at NLO, namely
\begin{eqnarray}
\Delta P_{ij}^{NLO}(z,t)&\equiv& \Delta P_{ij}^{(0)}(z)+\alpha_s(t) P_{ij}^{(1)}(z) \nonumber \\
P_{ij}^{NLO}(z,t)&\equiv& P_{ij}^{(0)}(z)+\alpha_s(t) P_{ij}^{(1)}(z).
\end{eqnarray}
involve a $t$ dependence associated with both the ${\bar MS}$ renormalization prescription for
$\alpha_s (t)$ 
\begin{equation}
\alpha_s(t)=e^{-(t-t_1)}\Bigl[1+b_1 (t-t_1)+\cdots \Bigr] \label{alphanlo}
\end{equation}
and with the factorization prescription used in the calculations. It is important to observe that
\begin{equation}
\Delta G^{NLO}(x,t)=\Delta G^{NLO}(x,t_1)+A_0^{\epsilon,NLO}(x,t_1)\cdot [G^{NLO}(x,t)-G^{NLO}(x,t_1)]+\mathcal{O}(t-t_1)^2, \label{DDGNLO}
\end{equation}
takes a form analogous to the LO result, except that the corrections $\mathrm{O} (t-t_1)^2$ to
(\ref{DDGNLO}) are necessarily smaller than those found in (\ref{DDG}). In (\ref{DDGNLO}),
$G^{NLO}(x,t)$ evolves in $t$ with the full NLO expression while evolution of $t$ in (\ref{alphanlo}) is 
only approximately correct at NLO. This is, however, an improvement on the LO result. Preliminary 
numerical calculations indicate that the same methods discussed in section 2.1 above can be used to 
determine fits of the form
\begin{equation}
A_0^{\epsilon,NLO}(x,t_1)\equiv A(t_1)x^{\alpha (t_1)}-(B(t_1)-1)x^{\beta (t_1)}+(B(t_1)-A(t_1))x^{\beta (t_1)+1},
\end{equation}
to parametrize the shape of $A_0^{\epsilon,NLO}(x,t_1)$.

We have not fully explored the range of boundary conditions defined by different parameterizations of
$\Delta g^{\epsilon}(x)$ for the NLO equations. For parameterizations consistent with (\ref{dgebnd}), we 
have found that the differences between extrapolations for $\Delta G(x,0)$ generated at LO and NLO 
occur primarily for $x\le 0.10$. It is expected, based on the above discussion, that errors associated with 
the extrapolation in $t$ of (\ref{DDG}) can be improved in NLO. This will be tested in the future. At 
present, however, we will address issues involving experimental data only at LO.

\section{Experimental Extraction of the Gluon Asymmetry and the Orbital Angular Momentum}

Section II describes an approach to the determination of $\Delta G(x,t)$ using the decomposition of 
Eq. (\ref{DGdef}) combined with experimental information on fits to $q(x,t)$, $G(x,t)$ and $\Delta q(x,t)$
to solve a nonlinear partial differential equation for $A_0^{\epsilon}(x)$. Since it uses the measured 
parton distributions in a unique way, this approach is complementary to determination of $\Delta G(x,t)$
based upon a ``global analysis" as exemplified in the fits of Hirai and Kumano \cite{acc} and as 
described in other analyses. \cite{bmn,bb}

In our approach, the function $\Delta g_{\epsilon}(x)$ describes a boundary condition on the partial
differential equation for $A_0^{\epsilon}(x)$ that represents nonperturbative dynamical constraints on
the DGLAP evolution of gluon helicity in a proton. The limits on $\Delta g_{\epsilon}(x)$ for which stable
solutions of Eq. (\ref{A0calc2}) can be found depend sensitively upon the parameterization of the input
parton distributions at the specific value of $t$ at which the separation 
$\Delta G(x,t)=A_0^{\epsilon}\cdot G(x,t)+\Delta g_{\epsilon}(x)$
is defined. We have explored those limits using the specific leading order (LO) fits to the input
distributions, Eqs. (\ref{CTEQ5})-(\ref{polq}) at $t=0$ ($Q^2= 1$ GeV$^2$). This tests the assumption that
DGLAP evolution can be combined meaningfully with the $J_z=\frac{1}{2}$ sum rule, Eq. (\ref{JSR}), at
this scale. The evolution of $\Delta G$ depends upon its initial shape, so its determination must be
accompanied by assumptions of large-$x$ behavior and interpolations with the small-$x$ shape. Our
approach to the shape provides strong theoretical constraints and differs from global fits \cite{acc} and
models such as the simpler valon model \cite{valon}, which make different assumptions about the
large-$x$ behavior of $\Delta G$. 

For those parameterizations of $\Delta g_{\epsilon}(x)$ for which a stable solution to
Eq. (\ref{A0calc2}) exists at $t=0$, we can insert the decomposition of Eq. (\ref{DGdef}) into 
Eq. (\ref{Leps}) to write
\begin{equation}
L_z(0)=0.34\pm 0.02-<A_0^{\epsilon}(x)\cdot G(x,0)>-<\Delta g_{\epsilon}(x)> \label{dLz}
\end{equation}
and evolve the equivalent possible forms of $\Delta G(x,0)$ to compare with experimental extractions of
$\Delta G(x,t)$ found through processes such as prompt photon and photon + jet measurements. 
The terms in Eq. (\ref{dLz}) use the experimental averages for $<\Delta \Sigma>$ from COMPASS and 
HERMES, and the CTEQ or MRST parameterizations for $G(x,t_0)$ from section II. The error quoted 
here is due to the uncertainties in the data, as previously mentioned and the small theoretical 
uncertainties associated with the unpolarized distributions. For those asymmetry parameterizations that 
best agree with data, we can extract the most likely range for the orbital angular momentum of the 
constituents.

From these results, a strictly theoretical approach to finding feasible values to $L_z$ yields 
a range allowed by the ``practical" constraints of
\begin{eqnarray}
&-0.12& \le <\Delta G> \le 0.62  \quad \mbox{and} \nonumber \\
&-0.28& \le L_z(0) \le 0.46,  \label{DGlim1}
\end{eqnarray}
where the theoretical and experimental uncertainties have been included in the quoted inequality 
values. The range is not sensitive to the unpolarized input distributions and the corresponding 
theoretical uncertainties have been included in this range. The ranges of $<\Delta G>$ and 
$<L_z>$ appear to be numerically limited by the constraints on $<\Delta g_{\epsilon}>$ in
equation (\ref{dgebnd}). From a theoretical point of view, this results in relatively small values of 
$\Delta G$ and $L_z$, but the range is still considerable. We must use the data on the asymmetry to 
further narrow the possible values of the gluon polarization and the orbital angular momentum. Table 3 
includes a summary of present data on the asymmetry [1-5].

\begin{table}[htdp]
\caption{Measurements of the gluon asymmetry}
\begin{center}
\begin{tabular}{|c|c|c|c|c|c|c|}
\hline
Experiment & Process & Value & stat & sys & $x_{avg}$ & Comment \\
\hline
\hline
COMPASS & open charm & -0.49 & $\pm 0.27$ & $\pm 0.11$ & 0.11 & Large errors \\
\hline
COMPASS & High $p_T$ hadrons & 0.016 & $\pm 0.058$ & $\pm 0.055$ & 0.09 & $Q^2 \ge$ 1 GeV$^2$ \\
\hline
COMPASS & High $p_T$ hadrons & -0.06 & $\pm 0.31$ & $\pm 0.06$ & 0.13 & $Q^2 \le$ 1 GeV$^2$ \\
\hline
SMC & High $p_T$ hadrons & -0.20 & $\pm 0.28$ & $\pm 0.10$ & 0.07 & $Q^2 \ge$ 1 GeV$^2$ \\
\hline
HERMES & Factorization method & 0.078 & $\pm 0.034$ & $\pm 0.011$ & 0.204 & Consistent with RHIC \\
\hline
HERMES & Approximation method & 0.071 & $\pm 0.034$ & $\pm 0.010$ & 0.222 & Consistent with RHIC \\
\hline
HERMES & High $p_T$ hadrons & 0.41 & $\pm 0.18$ & $\pm 0.03$ & 0.17 & $Q^2 \le$ 1 GeV$^2$ \\
\hline
\hline
\end{tabular}
\end{center}
\label{default}
\end{table}

It is virtually impossible to find parameterizations that completely agree with all data. Two separate 
analyses were performed: a $\chi^2$ calculation of all parameterizations with all existing data and a 
determination of which of the asymmetry parameterizations best agreed to within one to two $\sigma$ of 
the data set in Table 2. Six of the parameterizations in Table one, corresponding to lines 1, 2, 4, 6, 7 and 
9 all minimized $\chi^2$ and were within $2\sigma$ for at least five of 
the seven data points. All of the parameterizations that best  agree with data have the property that 
$\Delta g_{\epsilon}\le 0$.  A plots showing four of the parameterizations of $A(x,t_0)$ at LO are shown 
in figure 1. 

\begin{figure} 
\begin{center}
   \rotatebox{0}{\resizebox{4.0in}{!}{\includegraphics{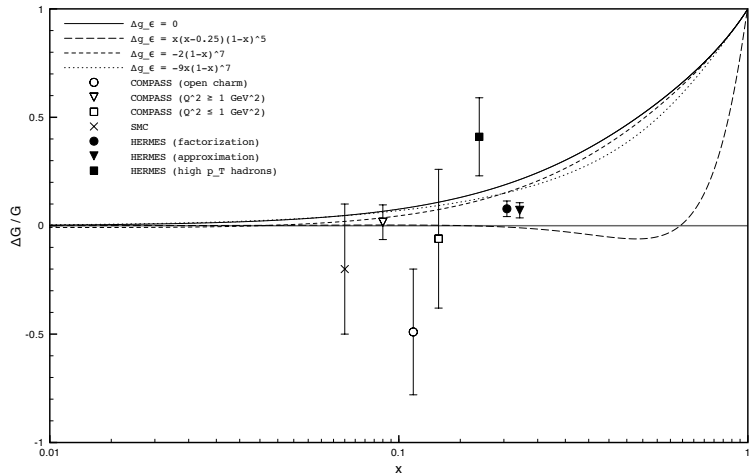}}} 
\end{center}   
\caption{The existing data for the polarized gluon assymmetry with the parameterizations that best agree with data.}
   \label{fig1}
\end{figure}

Using only these parameterizations, the range of $\Delta G$ and $L_z$ that result are
\begin{eqnarray}
&-0.12& \le \> <\Delta G> \> \le 0.31 \quad \mbox{and} \\
&0.03& \le \> L_z(0) \> \le 0.46, \label{DGlim2}
\end{eqnarray}
where, again, the uncertainties are included in the extreme values. These results are consistent with
data and the MIT Bag model, discussed by Chen and Ji. \cite{chenji}
Using present data, the range of
$\Delta G$ and $L_z$ have been narrowed, but clearly, accurate data over a wider kinematic range can 
more significantly constrain both the gluon polarization and the orbital angular momentum of the 
constituents. The results for the scenarios we have shown are not significantly different from each other. 
However, we do not rule out the necessity of altering this approach to consider more flexible 
parameterizations if data change significantly. We consider it likely that additional accurate data for the 
gluon asymmetry and $\Delta G$ over a wider kinematic range will allow us to further restrict our range
of consistency. Two immediate experimental possibilities are 
\begin{itemize}
\item (1) the measurement of $\Delta G/G$ in a wider kinematic range of $x_{Bj}$ for a fixed $Q_0^2$ value and 
\item (2) extraction of $\Delta G$ by prompt photon production or jet production in polarized $pp$ collisions (RHIC). 
\end{itemize}

Our results indicate that the integrated $<\Delta G>$ is likely positive and small at $Q_0^2\approx 1$
GeV$^2$. This is consistent with data \cite{hermes,compass3}, chiral quark models \cite{wakamatsu}
and the MIT bag model \cite{chenji}. Although most of our parameterizations of the asymmetry are
positive definite, the one in the second line of Table 2 changes sign and is consistent with data. This
possibility has been discussed by others and is not ruled out by present data \cite{chenji,acc,lss}. Many
of our parameterizations give a gluon polarization consistent with zero, in agreement with much of the
data from RHIC. \cite{star,PHENIX} Clearly there is still work to be done. However, we have provided
a mechanism for calculating the gluon asymmetry that allows extraction of information on both
$\Delta G$ and $L_z$. 

\begin{table}[htdp]
\caption{Comparison of Orbital Angular Momentum Models}
\begin{center}
\begin{tabular}{||c|c|c|c||}
\hline
Model & $L_z$ at $Q_0^2=1$ GeV$^2$ & Reference & Notes \\
\hline
\hline
CQSM & $0.25-0.32$ & \cite{wak} & $\mu_0^2=600$ MeV$^2$, no gluons \\
\hline
CQSM & $0.25$ & \cite{zavada} & $J_G\approx 0$, $L_q\le \frac{1}{3}$ \\
\hline
R$\chi$QM model & $0.25$ & \cite{zavada} & $\Delta G \le 0.10$ \\
\hline
$\chi$QM & $0.26$ & \cite{song} & no gluons, all $L_z^q$ \\
\hline
Skyrme & $0.50$ & \cite{song,Skyrme}  & no gluons \\
\hline
Casu-Seghal & $0.42$ & \cite{CaSe} &  \\
\hline
IK Quark model & $-0.25$ & \cite{BD} & $\Delta G = 0.59$, $L_z\approx L_q$ \\
\hline
MIT Bag & $0.18$ & \cite{chenji,song} & $\Delta G: 0.2-0.3$ \\
\hline
RGM+OGE+$\pi$ cloud & $0.35-0.61$ & \cite{AT} & $L_q$ contribution only \\
\hline
Quark model & $-0.43$ & \cite{MHS} & Maximal glue - GRSV \\
\hline
Quark model & $0.025$ & \cite{MHS} & Normal GRSV glue \\
\hline
Lattice calculation & $0.38$ & \cite{Lattice} & Unquenched: no glue \\
\hline
\end{tabular}
\end{center}
\label{default}
\end{table}

In Table 3, we outline some of the (mostly quark) models that make predictions about the orbital angular
momentum of the constituents (nearly all $L_z^q$ contributions). This includes various quark models 
combined with corresponding gluon scenarios. Most fall within the same range as our calculations,
except those predicting a large orbital angular momentum, either positive or negative. However, ours is 
a purely phenomenological approach that includes gluon contributions extracted from the asymmetry 
and does not rely entirely on present fits to data. We use the data only to constrain those models which
are most viable. Clearly, accurate data in a wider kinematic range of $x$ and $Q^2$ are necessary to
distinguish those models that are most viable and determine the total angular momentum of the
constituents. 

For illustrative purposes, we have shown a plot of the $x$-dependent function, $L_z(x,t=0)$ in figure 2 
for those parametrizations of $\Delta g_{\epsilon}$ included in figure 1. Note that they have different 
behavior at small-$x$, further motivating the need for accurate asymmetry data, especially at these 
values of $x$.

\begin{figure}
\begin{center}
   \rotatebox{0}{\resizebox{4.0in}{!}{\includegraphics{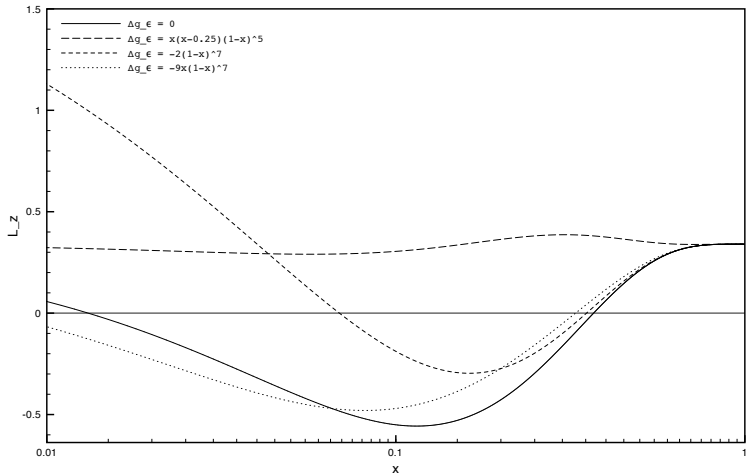}}}
\end{center} 
   \caption{Orbital angular momenta as a function of $x$ for parametrizations of the polarized gluon asymmetry that best agree with data.}
   \label{fig2}
\end{figure}

To summarize the relation between allowable values of $\Delta G$ and the orbital angular momentum of
constituents, figure 3 shows the range of $\Delta G$ values determined by our analysis and the
corresponding range of $L_z$ values, bounded by the vertical lines.
 
\begin{figure}
\begin{center}
   \rotatebox{0}{\resizebox{4.0in}{!}{\includegraphics{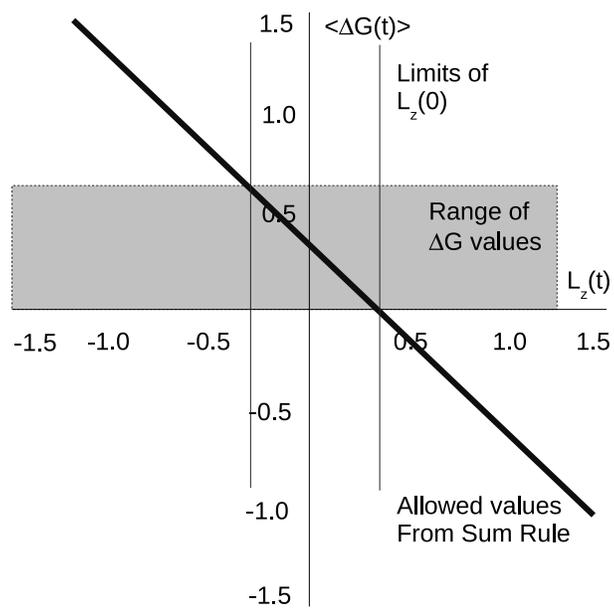}}} 
\end{center}
   \caption{Possible range of $\Delta G$ values as a function of the orbital angular momentum $L_z$.}
   \label{fig3}
\end{figure}

\section{Conclusion}

The property of color confinement implies that, at distance scales on the order of hadron sizes, the 
effective dynamic degrees of freedom for hadronic interactions are not given directly by perturbative 
interactions of quarks and gluons, but by some variety of collective excitations of the objects. Computer 
simulations of lattice regularized QCD have proven to be an effective tool for understanding many 
aspects of the crucial distance scale \cite{4.1}. In addition, the implications of the broken chiral symmetry 
of QCD associated with the light quark masses has been incorporated into a class of effective field 
theories for the pseudo-Goldstone bosons \cite{4.2} of the broken symmetry. This process has led to 
descriptions of nuclear structure such as the Georgi-Manohar chiral quark model \cite{mg} or versions of 
the chiral quark soliton model, originated by Wess and Zumino \cite{4.4} in the treatment of the Skyrme 
solution. \cite{Skyrme}

Our approach to the study of the gluon asymmetry presented in sections 2 and 3 has applied the 
$J_z=\frac{1}{2}$ rule and focussed on the ability to use information on $\Delta G(x,t)$ to estimate the 
orbital angular momentum $L_z(t)$ at low energy scales. This is due to the features of proton structure 
that can be described in terms of the constituent quark model involving non-relativistic ``constituent" 
quarks in an $s-$wave $L=0$ state. In this context, nonzero orbital angular momentum of colored 
constituents as inferred from equation (\ref{Leps}) is associated with the internal structure of the 
constituent quarks. The importance of such virtual corrections can also be studied in the flavor 
asymmetry of the $q\bar q$ sea, \cite{4.6} but the knowledge of $L_z(t)$ at values of $Q^2=1$ GeV$^2$ 
provides more complete information. For example, the orbital angular momentum of constituents in the 
Georgi-Manohar model has been studied by Song \cite{song} and Sivers \cite{4.8}. Song's results of 
$L_z=0.30$ are based upon a true effective field theoretical approach and does not consider the impact 
of gluons. The approach of reference \cite{4.8} estimates the impact of gluonic orbital angular 
momentum in Song's approach by requiring consistency between transverse and longitudinal spin. The 
results there are $L_z=0.39\pm 0.02$. 

We have described a decomposition of the scale-dependence of the gluon spin asymmetry,
$\Delta G(x,t)$. This approach allows a measurement sensitive to $\Delta G(x,t)$ to be extrapolated in
a manner consistent with DGLAP evolution, counting rules and the measured distributions
$\Delta q(x,t)$, $q(x,t)$ and $G(x,t)$ to help complete a more comprehensive dynamical picture of 
nucleon spin structure. The numerical work with the leading-order approximation to this equation 
indicates that the range of stability is acceptable. Furthermore, existing experimental measurements, 
within reported errors suggest that $\Delta G(x,t)$ lies within this range. To illustrate the value of this 
decomposition, we have considered the orbital angular momentum inferred from the $J_z={1\over 2}$ 
sum rule.

\end{document}